\documentclass[showpacs,twocolumn,aps,prl]{revtex4}
\usepackage{amsmath}
\usepackage{amsfonts}
\usepackage{amssymb}
\usepackage{amsthm}
\usepackage{graphicx}

\begin{document}

\title{Ratchet Effect in Magnetization Reversal of Stoner
Particles}

\author{Z. Z. Sun}
\affiliation{Physics Department, The Hong Kong University of
Science and Technology, Clear Water Bay, Hong Kong SAR, China}
\author{X. R. Wang}
\affiliation{Physics Department, The Hong Kong University of
Science and Technology, Clear Water Bay, Hong Kong SAR, China}
\date{\today}

\begin{abstract}
A new strategy is proposed aimed at substantially reducing the
minimal magnetization switching field for a Stoner particle.
Unlike the normal method of applying a static magnetic field which
must be larger than the magnetic anisotropy, a much weaker field,
proportional to the damping constant in the weak damping regime,
can be used to switch the magnetization from one state to another
if the field is along the motion of the magnetization.
The concept is to constantly supply energy to the particle from
the time-dependent magnetic field to allow the particle to climb
over the potential barrier between the initial and the target
states.
\end{abstract}
\maketitle
{\it Introduction--}The recent advance in technology allows the
fabrication of magnetic nano-particles\cite{Sun,Qikun} that
are potentially useful for high density information storage.
A magnetic nano-particle, in which the magnetic moments of all
atoms are aligned in the same direction, is called a Stoner
particle. Manipulation of a Stoner particle\cite{Hillebrands}
is of significant interest in information processing. Finding
an effective way to switch the magnetization from one state to
another requires a clear understanding of magnetization dynamics.
One important issue in magnetization reversal of Stoner particles
is the minimal switching field. This problem was first studied
by Stoner and Wohlfarth (SW)\cite{Stoner} who showed that
a field $h$ larger than the SW-limit $h_{SW}$ can switch the
magnetization from its initial state to the target value through
a ringing effect\cite{Hiebert,Acremann2,Crawford,He,xrw}.
However, recent theoretical and experimental studies\cite
{He,xrw,grenoble,Back,Schumacher} have shown that the minimal
switching field can be smaller than the SW-limit. Most studies
have assumed the magnetic field to be time-independent.
However, a very recent experiment\cite{grenoble} has shown that a
dramatic reduction of the minimal field is possible by applying a
small radio-frequency (RF) field pulse (the decrease in the
constant field is much larger than the amplitude of the RF-field).
In this study, it has been shown that a small time-dependent
magnetic field (ratchet) can affect the magnetization of a Stoner
particle such that the magnetization can move upward in its
energy landscape against the dissipation effect. A consequence
of this is that the minimal switching field is much smaller in
comparison with the case of a time-independent magnetic field.
In the case where the field magnitude does not change but the
direction is allowed to vary, it can also be shown that the
minimal field is proportional to the damping constant at the
weak damping limit.

\begin{figure}[htbp]
 \begin{center}
\includegraphics[width=8.5cm, height=4.cm]{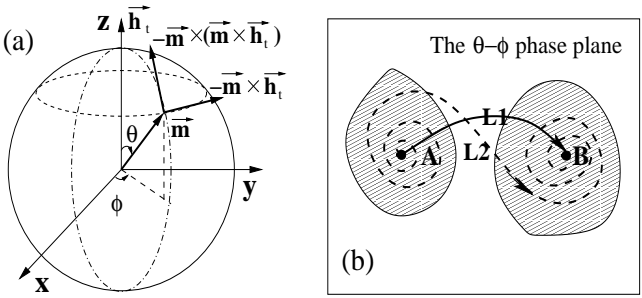}
 \end{center}
\caption{\label{fig1} (a) Two motions of magnetization $\vec{m}$:
$\vec{h}_t$ is the total magnetic field. $-\vec{m}\times
\vec{h}_t$ and $-\vec{m}\times (\vec{m}\times \vec{h}_t)$
describe the precession and dissipation motions, respectively.
(b) The $\theta-\phi$ phase plane. Two stable fixed points A
and B represent the initial and the target states, respectively.
Two shadowed areas denote schematically basins of A and B.
The solid curve L1 and dashed curve L2 illustrate two different
phase flows connected A and B. }
\end{figure}

{\it{Dynamics of magnetization in a magnetic field--}}The
magnetization $\vec{M}=\vec{m}M_s$ of a Stoner particle
can be conveniently described by a polar angle $\theta$
and an azimuthal angle $\phi$, shown in Fig.~\ref{fig1}(a)
where $\vec{m}$ is the unit direction of the magnetization,
and $M_s$ is the saturated magnetization of the particle.
In $\theta-\phi$ plane, each point corresponds to a
particular state of the magnetization. The evolution of a
state is governed by the Landau-Lifshitz-Gilbert (LLG)
equation\cite{xrw,Landau},
\begin{equation}
(1+\alpha^2)\frac{d\vec{m}}{dt} = - \vec{m} \times \vec{h}_{t}
- \alpha \vec{m} \times (\vec{m} \times \vec{h}_{t}), \label{LLG}
\end{equation}
where $t$ is in a unit of $(|\gamma|M_s)^{-1}$, and the
magnetic field is in the unit of $M_s$. $|\gamma|=2.21 \times
10^5 (rad/s)/(A/m)$ is the gyromagnetic ratio and $\alpha$ is
a phenomenological dimensionless damping constant. The typical
experimental values of $\alpha$\cite{Back} range from 0.037 to
0.22 for different Co films. The total field comes from an
applied magnetic field $\vec{h}$ and the internal field $\vec{h}
_i$ due to the magnetic anisotropy. Let $w(\vec{m},\vec{h}=0)$ be
the magnetic energy density function. Then $\vec{h}_{t}=-\nabla_
{\vec{m}} w(\vec{m},\vec{h})/\mu_0=\vec{h}_i+\vec{h}$ where $\mu_
0=4\pi \times 10^{-7}N/A^2$ is the vacuum magnetic permeability.
As shown in Fig.~\ref{fig1}(a), the first term in the right hand
side (RHS) of Eq.~\eqref{LLG} describes a precession motion
around the total field and the second term is the damping motion
toward the field.

The switching problem for a uniaxial Stoner particle is as
follows: Before applying an external magnetic field, there are
two stable fixed points (denoted by A and B in Fig.~\ref{fig1}(b))
corresponding to magnetizations, say $\vec{m}_0$ (point A) and
$-\vec{m}_0$ (point B) along its easy axis. The shadowed areas in
Fig.~\ref{fig1}(b) denote basins A and B. The system in basin A(B)
will end up at state A(B). Initially, the magnetization is
$\vec{m}_0$, and the goal is to reverse the magnetization to
$-\vec{m}_0$ by applying an external field as small as possible.
The issue is what is the minimal field $h_c$ defined as
$h_c=max\{h(t); \forall t\}$ for a given magnetic anisotropy.

{\it {Time-dependent vs. time-independent magnetic field--}}In
order to show that the magnetization reversal in a time-dependent
external magnetic field is qualitatively different from that in a
constant field, it is useful to look at the energy change rate.
From Eq. (\ref{LLG}),
\begin{equation}
\frac{dw}{dt} = -\frac{\alpha}{1+\alpha^2}(\vec{m}
\times \vec{h}_{t})^2-\vec{m}\cdot \dot{\vec{h}}, \label{energy}
\end{equation}
where $\dot{\vec{h}}$ is the time derivative of $\vec{h}$.
If the external field is time-independent, the second term on
the RHS vanishes, and hence the energy will always decrease.
In other words, a constant field is not an energy source.
Conversely, a time-dependent field can provide energy to
a particle. According to Eq.~\eqref{energy}, the second term
on the RHS can be either positive or negative depending on the
relative direction of $\vec{m}$ and $\dot{\vec{h}}$. This second
term can even be larger than the first one so that the particle
energy increases during its motion. $-\vec{m}\cdot
\dot{\vec{h}}$ is a maximum when $\vec{m}$ and $\dot{\vec{h}}$ are
in the opposite direction. From $|\vec{m}|=1$, it is known that
$\vec{m}$ and $\dot {\vec{m}}$ are orthogonal to each other,
which leads to $\vec {m}\cdot \ddot{\vec{m}}=-\dot{\vec{m}}\cdot
\dot{\vec{m}}$. The second term on the RHS of Eq. (2) is the
maximum when $\vec{h}=h_0\dot{\vec{m}}/|\dot{\vec{m}}|$ for a
fixed $h_0$. Then, from Eqs.~\eqref{LLG} and \eqref{energy}, the
maximal rate of energy increase is
\begin{equation}
\frac{dw}{dt} = \frac{|\vec{m} \times \vec{h}_{t}|}{\sqrt
{1+\alpha^2}}(h_0-\frac{\alpha}{\sqrt{1+\alpha^2}}|\vec{m}
\times \vec{h}_{t}|). \label{energy1}
\end{equation}
It should be highlighted that $\vec{h}$ is only well defined
when $\dot{\vec{m}}\ne 0$. Thus, in a numerical calculation,
some numerical difficulties will exist when the system is near
the extremes or the saddle points. Special care must be taken
at these points.

{\it {New strategy--}}A new strategy based on Eq.~\eqref{energy1},
can be developed using a smaller switching field . The field of
magnitude $h_0$ noncollinear with the magnetization was applied
to drive the system out of its initial minimum. Fluctuations
may also drive the system out of the minimum, but fluctuations
are inefficient. When the system is out of
the minimum and $\dot{\vec{m}}\ne 0$, a time-dependent field
$\vec{h}= h_0\dot{\vec{m}}/|\dot{\vec{m}}|$ is applied such that
$\dot{w}>0$. The system will climb the energy landscape
from the bottom. When the system energy is very close to the
saddle point, the field of magnitude $h_0$ can be rotated to
noncollinear with the magnetization, say $\pi/4$ to the direction
of the target state so that problems at $\dot{\vec{m}}=0$
are avoided and the system can move closer to the target state.
When the system has overcome the potential barrier between the
initial and target state and stays inside the basin of the
target state, the field can be turned off or applied in the
opposite direction to the motion of the magnetization, i.e.
$\vec{h}=-h_0\dot{\vec m}/|\dot{\vec m}|$. In the first case, the
system will reach the target state through the ringing motion caused
by the energy dissipation, often due to the spin-lattice relaxation.
In the second case, the system will move faster toward the target
state because both terms on the RHS of Eq.~\eqref{energy} will be
negative, resulting in a faster energy release from the particle.

\begin{figure}[htbp]
 \begin{center}
\includegraphics[width=7.5cm, height=4.8cm]{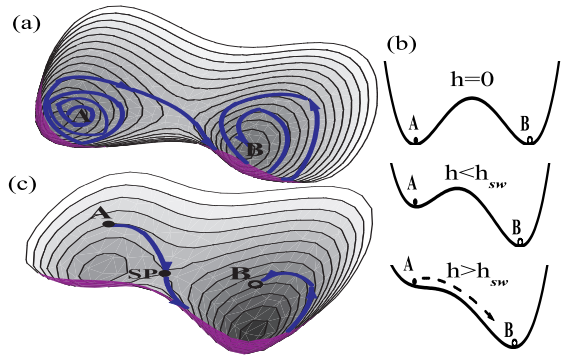}
 \end{center}
\caption{(a) Schematic illustration of the new strategy. A moving
(time-dependent) field acts as a ratchet for the magnetization.
The field along the motion of the magnetization provides the
energy to the particle initially in the left energy minimum such
that the system spins out of the minimum. Then a field opposite
to the motion of the magnetization causes the system to spin
into the right minimum.
(b) The SW-strategy: The target state (left minimum) becomes the
only minimum when the magnetic field is larger than $h_{SW}$.
The system will roll down along the potential landscape and
end up at the target state. (c) The strategy in the precessional
magnetization reversal: Both the initial (point A) and the
target (B) states are not local minima under the reversal field.
Due to the magnetization dynamics described by the LLG equation,
the particle will move along the trajectory denoted by the dash
line on the energy landscape. The field is switched off as soon
as the particle arrives B.\label{fig2} }
\end{figure}

The strategy is schematically illustrated in Fig.~\ref{fig2}(a).
The particle first spins out of its initial minimum by extracting
energy from the field, and then spins into the target state by
both energy dissipation and energy release (to the field).
Since the energy gain from the field is partially compensated by
the energy dissipation during the spinning-out process while both
the field and the damping consume energy in the spinning-in
motion, the particle moves out of its initial minimum slowly in
comparison with its motion toward the target state. It can be
readily seen in Fig.~\ref{fig2}(a) that the particle makes more
turns around the left minimum and fewer around the right minimum.
A similar result was experimentally confirmed in reference 10.
It will be shown later that a (linearly polarized) RF-field
used in reference 10 is not the optimum. In fact, a
circularly-polarized-like microwave (around 100GHz for a Co
film\cite{Back}) is enough to switch a magnetization. The new
strategy should be compared with those of the SW and the precessional
pico-second magnetization reversal\cite{Back,Schumacher}.
As illustrated in Fig.~\ref{fig2}(b) and \ref{fig2}(c), the SW
strategy is to apply a large enough field to destroy the minimum
initial state so that the particle can end up in the
target state (Fig.~\ref{fig2}(b)). In the precessional
magnetization reversal\cite{xrw}, the field is applied in such a
way that the energy of the initial state is larger than that at
the saddle point between the initial and final states.
When the particle moves down the energy landscape, it will
pass through the saddle point and arrive at the target state.
The magnetization switch is achieved if the field is switched
off at this point (Fig.~\ref{fig2}(c)).

For simplicity, consider the case of an uniaxial magnetic
anisotropy with the easy axis lying along the x-axis.
The general form of $w(\vec{m}, \vec{h})$ can be written as
\begin{equation}
w(\vec{m}, \vec{h}) =- \mu_0({1 \over 2}k m_x^2 - m_x h_x - m_y
h_y - m_z h_z),\label{form}
\end{equation}
where $h_x$, $h_y$, and $h_z$ are the applied magnetic fields
along x-, y- and z-axis, respectively. $k>0$ is the parameter
measuring the strength of the anisotropy.

\begin{figure}[htbp]
 \begin{center}
\includegraphics[width=7.0cm, height=4.0cm]{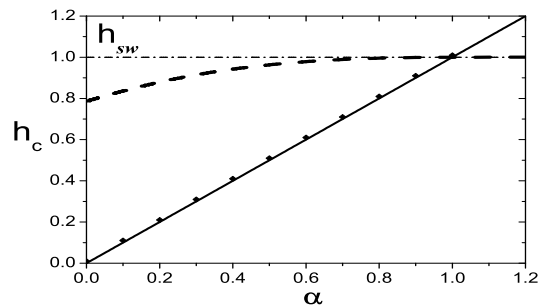}
 \end{center}
\caption{\label{fig3} The minimal reversal field (in unit $k/2$)
vs. the damping constant. The diamond symbols are the numerical
results of the new strategy for the uniaxial model of Eq. (4).
The solid curve is the analytical results. For comparison, the
dashed line is the minimal reversal field under a constant field
135$^{\circ}$ to the x-axis for the same magnetic anisotropy.}
\end{figure}

{\it {Results--}}To find the minimal switching field for the
uniaxial anisotropy of Eq.~\eqref{form}, it can be seen from
Eq.~\eqref{LLG}, that $\dot{\vec{m }}$ is linear in the magnetic
field, and as illustrated in Fig.~\ref{fig1}(a), each field
generates two motions for $\vec{m}$. The first is a precession
around the field, and the second toward the field. Under the
influence of the internal field (along the x-axis) and of the
applied field $\vec{h}=h_0\dot{\vec{m}}/|\dot{\vec{m}}|$, the
system evolves into a steady precession state for a small $h_0$
because the precession motion due to the applied field can
exactly cancel the damping motion due to the internal field.
The net motion (sum of precession around the internal field
and damping motion due to the applied field) is a precession
around the x-axis (easy axis). In this motion, the energy loss
due to damping and the energy gain from the time-dependent
external field are equal. The balance equation is
\begin{equation}
h_0-k\alpha \cos\eta \sin \eta =0, \label{steady}
\end{equation}
where $\eta$ is the angle between the magnetization and the
x-axis. The initial state is around $\eta=0$, any stable
precession motion must be destroyed in order to push the system
over the saddle point at $\eta=\pi/2$. Since Eq.~\eqref{steady}
has solutions only for $h_0 \le \alpha k/2$, the critical field is
\begin{equation}
h_c=\alpha k/2.
\label{field}
\end{equation}
It is of interest to note that the minimal reversal field is
proportional to the damping constant, and approach zero when
the damping constant goes to zero irrespective of how large the
magnetic anisotropy. For an arbitrary magnetic anisotropy,
it may not be possible to find the analytical expression for the
minimal reversal field, and should thus use numerical calculations.
To demonstrate that this can indeed be done numerically,
a calculation for the magnetic anisotropy of Eq.~\eqref{form} has
been performed. The result of the minimal reversal field vs.
damping constant $\alpha$ is plotted in Fig.~\ref{fig3}.
For comparison, the minimal reversal field for a time-independent
magnetic field laying at 135$^{\circ}$ from the x-axis has be plotted.
As it was explained in reference \cite{xrw}, the minimal reversal
field is smaller than the SW-limit for a small damping constant
$\alpha<\alpha_c$ (which is 1 for the model given by Eq.~
\eqref{form}) and equals to the SW-limit for $\alpha>\alpha_c$.
It is clear that the new strategy is superior to that of SW or
precessional reversal scheme only for $\alpha<1$, and is worse
for larger $\alpha$.

\begin{figure}[htbp]
 \begin{center}
\includegraphics[width=6.5cm, height=3.0cm]{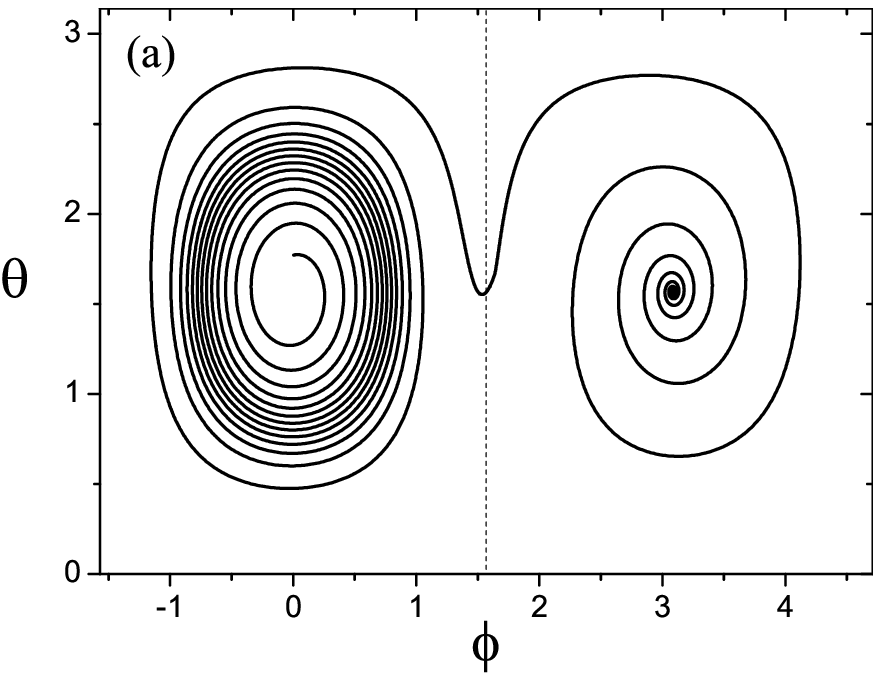}
\includegraphics[width=7.cm, height=5.cm]{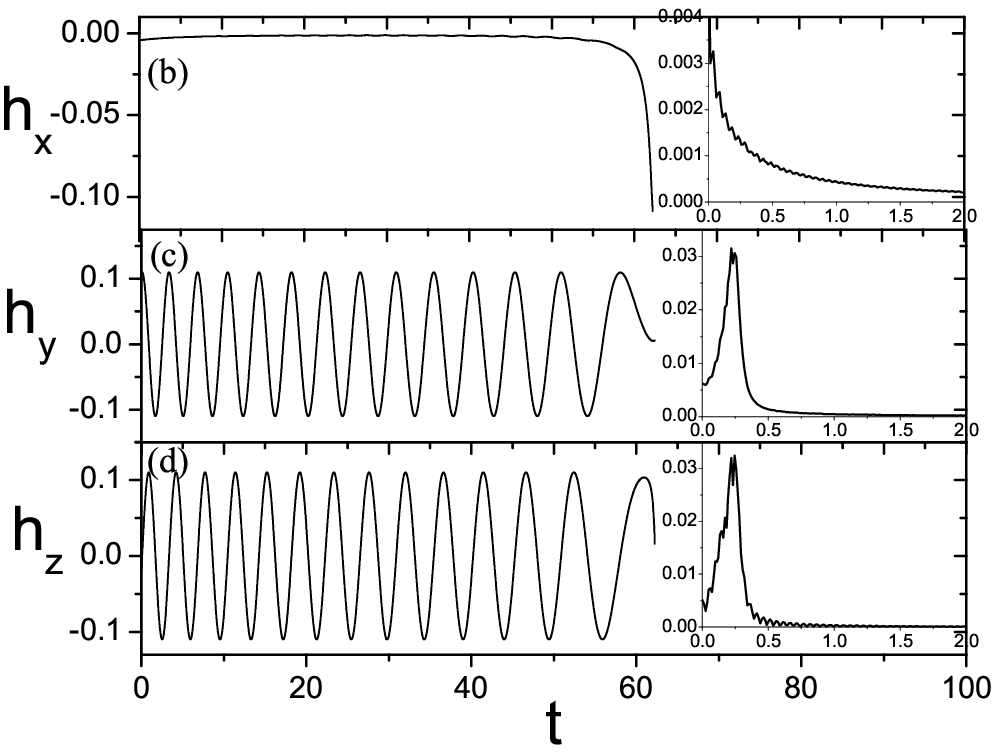}
 \end{center}
\caption{\label{fig4} (a) The phase flow under the new strategy
with $k=2$, $\alpha=0.1$ and $h_0=0.11$. Just as illustrated in
Fig. 2(a), the phase flow shows a slow spin-out motion near the
initial state and a fast spin-in motion near the target state.
(b)-(d) The time-dependent reversal field with the same
parameter as that in (a). Insets: The corresponding Fourier
transforms.}
\end{figure}

To explain the type of field to be used in this new strategy,
the trajectory of the system is numerically calculated and
the time-dependent magnetic field is recorded.
The results for $k=2$, $\alpha=0.1$, and $h_0=0.11>h_c$ are given
in Fig.~\ref{fig4}. Fig.~\ref{fig4}(a) is the phase flow of the
system starting from a point very close to the left minimum.
As explained previously, the particle moves many turns
in the left half of the phase plane before it crosses the
potential barrier (the saddle point is on the middle line) while
it moves toward the right minimum (the target state) much faster
(with few turns). Fig.~\ref{fig4}(b)-(d) are the corresponding
time dependence of x-, y-, and z-components of the magnetic field.
From these curves, it can be shown that $h_y$ and $h_z$ oscillate
with time reflecting the spinning motion around minima. In general
the spinning periods along different paths vary. Thus the
time-dependent magnetic field contains many different frequencies
as can be seen from the Fourier transform of $h_i(t), \
i=x,y,z$ shown in the insets of Fig.~\ref{fig4}(b)-(d).
For Co-film parameters of $M_s=1.36\times10^6 A/m$\cite{Back},
the time unit is approximately $(|\gamma|M_s)^{-1}=3.33ps$.
Correspondingly, the field consists of circularly-polarized
microwaves of about $100GHz$.

{\it {Discussion and conclusions--}}It should be noted
that the ratchet effect has already been used in many different
fields in condensed matter physics, including the manipulation
and control of vortex motion in superconducting films\cite{nori}
and smooth epitaxial film growth\cite{barabasi}. Although the
switching field in the new scheme is much smaller than that in
the old ones, it is an experimental challenge to create a
time-dependent magnetic field required by the new strategy.
A device that is sensitive to the motion of the magnetization
may be needed such that a coil can be attached to generate
the required field. It should be emphasized that the results
are based on the LLG equation which does not include
any quantum effects. Quantum effects may be important for small
particles whose level spacings are comparable with the energy
quanta of the time-dependent field. In that case, a quantum
version of LLG equation needs to be developed, which is beyond
the scope of the present work.

In conclusion, a scheme is proposed to dramatically reduce
the magnetization reversal field based on the fact that a
time-dependent magnetic field can be both energy source and
energy sink, depending on whether the field is parallel or
anti-parallel to the motion of the magnetization. The idea is
to constantly supply energy to a Stoner particles from the
time-dependent magnetic field to allow the particle to move out
of its initial minimum and to climb over the potential barrier.
After the particle lands in the basin of the target state,
the time-dependent field will act as an energy sink that
constantly withdraw energy from the particle such that the
particle will accelerate to the target state.
In a simple model with an uniaxial magnetic anisotropy,
the conditions and the solution of the steady precession
motion in the new scheme for $h_{0}<h_c$ were also found.

{\it{Acknowledgments}--}This work is supported by UGC, Hong Kong,
through RGC CERG grants. XRW would like to thank the hospitality
of Laboratoire Pierre Aigrain, Ecole Normale Supemrieure.
Discussion with Dr. G. Bastard, Prof. P. Tong, and Dr. Ke Xia is
acknowledged.

\end{document}